\documentclass[authoryear,sort&compress,nopreprintline,review,12pt]{elsarticle}

\usepackage{siunitx}
\usepackage{amssymb,amsmath}
\usepackage[shortcuts]{extdash} 
\usepackage[colorlinks=true,citecolor=blue]{hyperref}
\usepackage[version=4]{mhchem}
\ExplSyntaxOn
\keys_define:nn { mhchem }
{
	arrow-min-length .code:n =
	\cs_set:Npn \__mhchem_arrow_options_minLength:n { {#1} } 
}
\ExplSyntaxOff

\usepackage{siunitx,tabularx,multirow}
\usepackage[dvipsnames]{xcolor}

\usepackage[font=footnotesize,labelfont=bf,labelsep=period]{caption}   
\usepackage[font=footnotesize,labelfont=bf,labelsep=period]{subcaption}                     
\DeclareCaptionSubType*[arabic]{figure}                                                     
\DeclareCaptionLabelFormat{subfiglabel}{Figure #2}                                          
\usepackage{placeins}

\makeatletter
\def\ps@pprintTitle{%
	\let\@oddhead\@empty
	\let\@evenhead\@empty
	\def\@oddfoot{}%
	\let\@evenfoot\@oddfoot}
\makeatother

\graphicspath{{figures/}}

\DeclareSIUnit{\thomson}{Th}

\newcommand\mycitep[1]{(\citetalias{#1}, \citeyear{#1})}
\defcitealias{USDA2022}{USDA}
\defcitealias{InternationalAgencyForResearchOnCancer2010}{WHO}

\begin{document}

\begin{frontmatter}

\title{Effect of Burn Parameters on PAH Emissions at Conditions Relevant for Prescribed Fires}

\author[Stanford]{Karl T{\"o}pperwien\corref{cor1}}
\cortext[cor1]{Corresponding author. Department of Mechanical Engineering, Stanford University, 440 Escondido Mall, Stanford, California 94305, United States.}
\ead{karlt@stanford.edu}
\affiliation[Stanford]{
	organization={Department of Mechanical Engineering, Stanford University}, 
	addressline={440 Escondido Mall},
	city={Stanford}, 
	postcode={94305},
	state={California}, 
	country={United States}}
\author[Stanford]{Guillaume Vignat}
\author[PULSE]{Alexandra J. Feinberg}
\affiliation[PULSE]{
	organization={Stanford PULSE Institute, Department of Applied Physics, Stanford University}, 
	city={Stanford},
	postcode={94305},
	state={California}, 
	country={United States}}
\author[Aerodyne]{Conner Daube}
\affiliation[Aerodyne]{
	organization={Aerodyne Research Inc.}, 
	city={Billerica},
	postcode={01821},
	state={Massachusetts}, 
	country={United States}}
\author[Aerodyne]{Mitchell W. Alton} 
\author[Aerodyne]{Edward C. Fortner}
\author[Aerodyne]{Manjula R. Canagaratna}
\author[PULSE,SLAC]{Matthias F. Kling}
\affiliation[SLAC]{
	organization={Department of Photon Science, SLAC National Accelerator Laboratory}, city={Menlo Park},
	postcode={94025}, 
	state={California},
	country={United States}}
\author[StanfordMed,Harvard]{Mary Johnson}
\affiliation[StanfordMed]{
	organization={Sean N. Parker Center for Allergy \& Asthma Research}, 
	city={Palo Alto},
	postcode={94304},
	state={California} , 
	country={United States}}
\affiliation[Harvard]{
	organization={Department of Environmental Health, Harvard T.H. Chan School of Public Health}, 
	city={Boston},
	postcode={02115},
	state={Massachusetts}, 
	country={United States}}
\author[StanfordMed,Harvard]{Kari Nadeau}
\author[Aerodyne]{Scott Herndon}
\author[Aerodyne]{John T. Jayne}
\author[Stanford,SLAC]{Matthias Ihme}

\begin{abstract}
Wildfire smoke is a health hazard as it contains a mixture of carcinogenic volatile compounds and fine particulate matter. 
In particular, exposure to polycyclic aromatic hydrocarbons (PAHs) is a major concern, since these compounds have been recognized as important contributors to the overall carcinogenic risk of smoke exposure. 
In this work, gas and particle-phase PAH emissions from the combustion of Eastern White Pine (\emph{pinus strobus}) were quantified using time-of-flight mass spectrometry over a range of burn conditions representative of wildfires and prescribed fires.
These experiments allow for controlling conditions of fuel moisture, heat flux, and oxygen concentration to understand their impact on PAH emissions.
We find that optimal conditions for fuel moisture content of $20 - 30\%$, heat load onto the sample of $ 60 - \SI{70}{\kilo\watt\per\metre\squared}$, and oxygen concentrations of the burn environment of $5 - 15\%$ can  reduce the emissions of the heavy molar weight PAHs by up to 77\%.
Our analysis shows that the relative carcinogenic risk can be reduced by more than 50\% under optimal conditions, offering a way for reducing emission exposure from forest treatment activities.

\end{abstract}
	
\end{frontmatter}

\section{Introduction}
With climate change as a major driving factor \citep{Westerling2006}, wildfires have increased in severity and frequency over the past decades \citep{Spracklen2009} and are considered a major public health issue \citep{Gould2024}.
Growing evidence suggests that extended periods of exposure to wildfire smoke cause respiratory morbidity and increased mortality \citep{Reid2016,Gould2024}.
Volatile organic compounds (VOCs) are among the primary pollutants in wildfire smoke \citep{Karl2007,Yokelson2013,Koss2018,Sekimoto2018,Jaffe2020,Majluf2022}.
Polycyclic aromatic hydrocarbons (PAHs) constitute a subset of organic toxins with links to cancer \citep{Bostrom2002}, respiratory morbidity \citep{Godish2004}, and suppressed immune functions~\citep{Liu2013,Hew2015}.
PAHs are formed during incomplete combustion and pyrolysis of biomass, and may be present in both the gas \citep{Samburova2016} and particle-phase \citep{Eriksson2014,Jen2019}.

To mitigate the risk of severe wildfires, efforts to reduce excessive fuel buildup have been intensified in recent years \citep{Melvin2020}. 
Among the main tools for this important task are controlled and low-intensity prescribed fires, i.e., deliberate ignition of fuel to serve forest management objectives.
Beyond their potential benefits to ecosystem health \citep{Kalies2016}, they constitute an effective approach to reduce fuel loads, specifically of surface fuels \citep{Agee2005}.
This forest management practice also aims at reducing the intensity and spread rate of potential future wildfires in order to facilitate their suppression \citep{Hunter2020}.

Prescribed fire treatments may also be viewed as a tool to mitigate  harmful smoke exposure from wildfires \citep{Kiely2024}, but, in that regard, their overall effectiveness and benefits to human health and air pollution remain inconclusive \citep{Jones2022}.
Despite this knowledge gap, prescribed fire activity is growing considerably:
in the United States, the U.S. Forest Service has announced a 10-year plan to treat up to an additional 8 million hectares of National Forest System land by 2032 \mycitep{USDA2022}. 
This continues a past effort which saw an increase of the total area subjected to prescribed fire treatment from 1 to over 2.5 million hectares from 2010 -- 2018, albeit with substantial variations by region, in part due to public concerns over potential negative health outcomes associated with air pollution and smoke exposure \citep{Kolden2019}.
Therefore, to support these scaling efforts, there is an urgent need for better understanding the health impacts of such forest management practices, which are not well documented in the literature \citep{Williamson2016,Prunicki2019,Jaffe2020,Schollaert2023}.

Primary emissions from wildfires and prescribed fires depend on factors such as fuel type and conditions, fuel consumption rates, topography, and weather.
An advantaged of prescribed burns is that some of these factors (\emph{e.g.,} fuel conditions, consumption rates, fuel moisture\ldots) can be manipulated, controlled, or deliberately selected to some extent, which may result in reduced air pollution compared to wildfires \citep{Liu2017b}.
More specifically, portable devices such as incinerators offer extensive control over the combustion process compared to larger broadcast burns \citep{Lee2017}, and may be utilized for forest management in areas where tight control over pollutant emissions is critical, such as in the vicinity of densely populated areas.
These devices allow for a modulation of air flow rates, which has been observed to govern heat transfer, combustion efficiency, and consumption rates, alongside other parameters including fuel type and moisture content \citep{Porteiro2010}.
Pre-burn fuel manipulation is an alternative approach to influence fire spread rates or fire line intensities during broadcast burns, which can be beneficial to dispose of thinning residue and thereby reduce the fuel bed depth when incinerators cannot be used \citep{Pique2018}.

The formation and growth of PAHs is affected by many factors at different scales, with precursors typically formed during pyrolysis \citep{Richter2000}.
Increased aromatization is observed with increasing temperature, promoting the formation of benzene and naphthalene depending on the relative abundances of biopolymers in a given fuel sample \citep{Sekimoto2018}.
Subsequent heat addition by exothermic combustion can provide sufficient energy for cyclization reactions \citep{Frenklach1987} to form heavier PAHs.
However, above a certain temperature, PAH fragmentation becomes the dominant process \citep{Saggese2013}.
Partitioning between the gas and particle-phase correlates with the vapor pressure of a given compound, resulting in smaller PAHs being preferentially emitted into the gas-phase, while heavier PAHs tend to condensate onto the particle phase \citep{Lima2005}.
Different oxidation pathways in the atmosphere further affect the concentrations and lifetime of PAHs \citep{ODell2020}, which may lead to the formation of even more harmful compounds \citep{Albinet2008}.

Laboratory-scale experiments \citep{Jen2019}, pile burns \citep{Aurell2017}, and sampling of wildfire smoke plumes \citep{Permar2021} have underscored a prevailing negative correlation between PAH emissions and modified combustion efficiency (MCE), highlighting the influence of burning regimes (flaming vs. smoldering combustion) \citep{Gilman2015,Kortelainen2018,Koss2018}.
Beyond MCE, fuel moisture content (FMC) directly affects a sample's temperature evolution and dry time \citep{Blasi2003}, and is positively correlated with PAH emissions from biomass combustion \citep{Shen2013}.
The fuel type also constitutes an important parameter governing PAH emissions \citep{Jenkins1996}.

Despite these efforts, substantial differences persist in reported PAH emission factors, even when considering a single fuel type \citep{Yokelson2013,Prichard2020}.
As an example, the SERA database \citep{Prichard2020}, which compiles emission factors from a wide range of literature sources, reports the emission factor for phenanthrene as $\SI{3.2}{\milli\gram\per\kilo\gram}$, with a standard deviation of 69\,\% across published studies which utilized laboratory burns to quantify these emissions.  
This is in part due to complications in precisely controlling and characterizing the burn environment. 
Another difficulty is controlling for PAH oxidation in the atmosphere, which affects measurements differently depending on the burn configuration of the experiment \citep{Lima2005,Aurell2017,ODell2020,JuncosaCalahorrano2021}.
These parameters add to the uncertainties associated with assessing potential health outcomes from prescribed fires.

Our objective is thus to isolate the effect of individual parameters of the burn environment on PAH emissions,
and investigate the extent to which a more optimal control of the local burn environment could offer a pathway for reducing PAH emissions in prescribed fires.
We demonstrate that an optimal burn environment contributes to reducing negative health impacts from PAH exposure and is thus conducive to improving strategies for prescribed burns.
We aim at reducing emissions of higher-weight PAHs identified among the U.S. Environmental Protection Agency's (EPA) priority pollutant PAHs \citep{Yan2004}, as carcinogenic risk tends to increase with the mass of a given PAH \citep{Bostrom2002}.

Unlike prior studies, we precisely control the heat load and oxygen content of the combustion environment, as well as the FMC in a laboratory-scale combustor.
The investigated conditions represent a subset of all possible burn conditions that are likely to occur in real prescribed fires.
These parameters were intentionally chosen to create repeatable burn environments, enabling a clearer understanding of how altering a single parameter influences the resulting PAH concentration profiles.
Additional factors, such as the species of biomass, can influence PAH emissions, but are not considered in the present work as they have been studied elsewhere \citep{Jenkins1996,Koval2022}.
Considering both gas-phase and particle-phase data, we highlight a specific burn parameter range which minimizes overall PAH emissions.
Ultimately, we discuss implications for prescribed fire activities in the wild, including their assessment in terms of overall benefits and potential health risks, as well as limitations of the present approach.

\section{Materials and methods}
\label{sec:methods}

\subsection{Biomass samples}
Combustion of Eastern White Pine samples (\emph{pinus strobus})\---a representative fuel for the Eastern U.S.\---was conducted in a tunable combustion chamber (adapted from previous work by \cite{Boigne2021a}) at Aerodyne Research, Billerica, MA, USA. 
This fuel type will likely be affected by increased forest treatment activities, including prescribed fires, and shows comparably high PAH emissions among Eastern American fuels \citep{Fine2001}.

\begin{figure*}
	\centering
		\includegraphics[width=\textwidth,trim={2cm 9cm 2cm 0},clip]{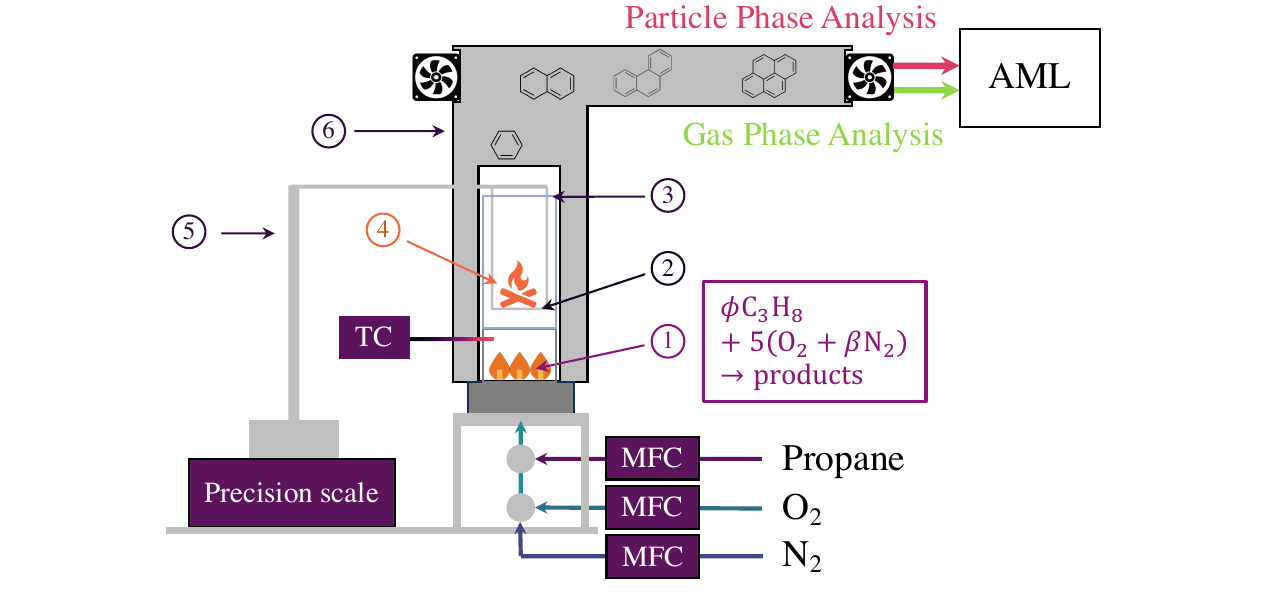}
	\caption{Illustration of experimental setup:
	(1): premixed propane/air burner; (2): sample holder (stainless steel wire screen); (3): transparent quartz tube; (4): \emph{pinus strobus} sample on sample holder; (5): sample holder post; (6): exhaust stack.
	AML: Aerodyne Mobile Lab; TC: Thermocouple; MFC: Mass flow controller. 
	}
	\label{fig:big_picture}
\end{figure*}

\subsection{Experimental setup}

To characterize the burn emissions, the combustion chamber was interfaced with the Aerodyne Mobile Laboratory (AML, see Fig.~\ref{fig:big_picture}), the details of which have been described elsewhere~\citep{Kolb2004,Herndon2005}. 
Briefly, the AML is a truck outfitted with atmospheric and aerosol spectrometers, including a proton-transfer time-of-flight mass spectrometer (PTR-ToF-MS, henceforth Vocus) to analyze VOCs in the gas phase.
Particle-phase aerosols are characterized by a High-Resolution Time-of-Flight Soot-Particle Aerosol Mass Spectrometer (HR-ToF-SP-AMS \citep{DeCarlo2006,Canagaratna2007}).

Unlike other experiments that investigate emissions from multiple fuel types at fixed burn conditions, we hold the fuel type constant to observe the effect of the burn environment on PAH emissions.
In suspending individual biomass samples (labeled as (4) in Fig.~\ref{fig:big_picture}) in a homogeneous stream of oxygen-deprived gases originating from a tunable premixed \ce{C3H8}/\ce{O_2}/\ce{N_2} matrix flame (1), we have direct control over heat flux and $\ce{O2}$-depletion of the flow around the sample, emulating fire conditions in a well-controlled and repeatable burn environment.
This is achieved by independently adjusting the equivalence ratio $\phi$ and the oxygen dilution ratio $\beta$ of the premixed matrix flame to yield the desired heat flux and $\ce{O2}$-depletion.
The flow rates of \ce{C_3H_8}, \ce{O_2}, and \ce{N_2} were controlled independently.

Additionally, fuel moisture content was modulated prior to combustion by pre-drying the samples in an oven.
These three parameters constitute the burn environment that was selected through high-coverage sampling of the three-dimensional parameter space. 
The specific heat flux from the oxygen-deprived gas stream of \ce{C_3H_8}/\ce{O_2}/\ce{N_2} combustion onto the sample was varied between $\dot{q}^{\prime\prime} = 33 - \SI{80}{\kilo\watt\per\meter\squared}$, in a range representative of realistic fires \citep{Silvani2009,Cruz2011,Frankman2013}.
The oxygen concentration was varied between $X_{\ce{O2}} = 1 - 23.5 \%$, and fuel moisture content (FMC) was varied between $\mathrm{FMC} = 1 - 58\%$ (based on the undried sample weight).
The flow velocity $v = \SI{2.1}{\metre/\second}$ around the sample was held constant.
A total of 23 samples were burned (sample length: $l = \SI{63.5}{\milli\metre}$, diameter: $d = 9 - \SI{34}{\milli\metre}$).
The sample size was dictated by the burner diameter ($d_{\mathrm{burner}} = \SI{70}{\milli\metre}$) and the weight limit of the scale.
Each burn took approximately 5 -- 10 minutes to proceed to completion, which was defined as the state at which the measured mass loss rate reached zero.
Because of the high sensitivity of the utilized instruments for gas and particle-phase characterization, the exhaust stream was diluted with ambient air as it passed through an exhaust pipe ($d = \SI{20}{\centi\metre}$) from which emissions were sampled by the AML.

\subsection{Instrumentation}
\subsubsection{Gas-phase analysis}
The Vocus employed a high-resolution ToF mass analyzer (Aerodyne Research Inc., USA, and TofWerk, Switzerland) at a mass resolution of greater than $9500$ FWHM $m/\Delta m$, and a temporal resolution of $\SI{1}{\hertz}$.
The time-resolved mass spectra were analyzed using Tofware v3.3.0 \citep{Stark2015} with IgorPro v9.0.2.4.
Automated calibrations were performed every $\SI{11}{\minute}$ with a reference mixture of VOCs taken from a calibration tank (Apel-Riemer Environmental Inc, Broomfield, CO), while instrument backgrounds were taken by overflowing the Vocus inlet with ultra-zero air.
Calibration factors for molecules in the calibration tank were inferred through correlation of their concentrations to ion intensities of the calibration periods.
In turn, calibration factors for the PAHs were estimated using a correlation with their respective reaction rate constant with $\ce{H3O+}$ \citep{Zhao2004,Sekimoto2017}, and subsequent normalization to the calibration factor of $\ce{(C6H6)H+}$.
An example of the fully constrained peak fit is shown in Fig.~\ref{fig:peak_fit_igor} for $\ce{(C14H10)H+}$, which we have tentatively identified as phenanthrene/anthracene (PH/AN).

For each burn condition, corresponding background measurements were taken with the \ce{C_3H_8}/\ce{O_2}/\ce{N_2} flame ignited, but without a biomass sample.
PAH emissions from the \ce{C_3H_8}/\ce{O_2}/\ce{N_2} were at least one order of magnitude lower than emissions during biomass combustion, and deemed negligible. 
The modified combustion efficiency \citep{Ward1993}, $\mathrm{MCE} = \Delta \ce{CO2}/(\Delta \ce{CO2} + \Delta \ce{CO})$, was obtained from background-subtracted measurements via tunable infrared laser differential absorption spectroscopy (TILDAS) \citep{McManus2011}.
Finally, we computed burn-averaged concentrations for each PAH to investigate how a changing burn environment affects the mean emission of a given PAH.
For three characteristic burn conditions (discussed later), we also present a particle-phase analysis.
Due to varying dilution of the burn emissions with ambient air, we report PAH concentrations normalized to benzene to compare emissions from different burns (see Sec.~\ref{ssec:gas_phase results}).

\subsubsection{Particle-phase analysis}
An Aerodyne Soot Particle Aerosol Mass Spectrometer (SP-AMS) measured the size ($\SI{100}{\nano\metre} - \SI{2.5}{\mu\metre}$) and bulk chemical composition of sub-micrometer non-refractory particulate matter and black carbon.
Details of the SP-AMS are described elsewhere \citep{Lee2015,Onasch2012,Willis2016}.
Briefly, aerosols are sampled with a volumetric flow rate of $\SI{85}{\centi\metre^3 \second^{-1}}$ and focused into a narrow beam by an aerodynamic particle focusing lens. 
Particle sizing was determined as particles pass a chopper when the AMS was in particle time-of-flight (PToF) mode. 
Non-refractory organics, nitrate, sulfate, ammonium, and chloride were vaporized by surface impaction on a heater at $\SI{873}{\kelvin}$.
Refractory black carbon-containing particles were heated and vaporized by passing through a $\SI{1064}{\nano\metre}$ laser beam reaching temperatures exceeding $\SI{4000}{\kelvin}$. 
Once vaporized, resulting gas phase molecules were ionized with 70 eV electron impact ionization. Ions formed were detected by a time-of-flight mass spectrometer. 
The AMS was typically run in Fast Mass Spec (FMS) mode, producing $\SI{1}{\second}$ time-resolved mass concentrations ($\SI{}{\mu\gram/\metre^{3}}$) for organics, nitrate, sulfate, ammonium, chloride, and black carbon. 
The SP-AMS was operated in FMS primarily with a PToF sizing mode interval occurring occasionally. 
A PM$_{2.5}$ inlet was installed on this SP-AMS allowing for the detection of larger particles relative to the standard lens typically used with the AMS. 
The inlet from the front of the AML to the SP-AMS had a flow rate of 5 LPM with a residence time in the inlet of approximately $\SI{6}{\second}$.

\subsection{Computation of heat fluxes onto the samples}
\label{ssec:heat_flux_calculation}
The heat flux from the oxygen-deprived hot gas stream onto the sample is computed as a ``gauge heat flux'' (the amount of energy that would be absorbed if the sample was cold):
\begin{equation}
	\dot{q}^{\prime\prime} = h (T_{gas} - T_{0}),
	\label{eqn:heat_flux}
\end{equation}
where $T_{gas}$ denotes the temperature of the gas stream resulting from \ce{C_3H_8}/\ce{O_2}/ \ce{N_2} combustion, and $T_{0}$ the initial temperature of the sample, assumed to be $T_{0} = \SI{300}{\kelvin}$.
The heat transfer coefficient $h$ is determined as
\begin{equation}
	h = \frac{\mathrm{Nu} \cdot k}{d}
	\label{eqn:heat_transfer_coeff}
\end{equation}
using the thermal conductivity $k$ of the gas and the diameter $d$ of the sample (assumed to be cylindrical).
The Nusselt number, Nu, is obtained from a standard correlation \citep{bergman2011fundamentals} for cylindrical objects as a function of the flow Reynolds number Re inside the combustion chamber and the Prandtl number Pr of the gas stream.
We compute bulk gas properties with Cantera \citep{cantera} using the GRI3.0 mechanism \citep{GRI30} for each burn condition to determine the gas stream's density $\rho$, its viscosity $\mu$, Pr, and temperature $T_{gas}$ at equilibrium.
With the flow velocity defined as $v = \dot{m}/(\rho  S)$ we evaluate the flow Reynolds number $\mathrm{Re} = \rho v d/\mu$, and ultimately obtain the Nusselt number as \citep{bergman2011fundamentals}
\begin{equation}
	\mathrm{Nu} = 0.638 \cdot \mathrm{Re}^{0.466} \cdot \mathrm{Pr}^{1/3}.
\end{equation}
Note that the mass flux $\dot{m} = \SI{3}{\kilo\gram\per\metre\squared\per\second}$ and the cross-sectional area $S$ where held constant for all conditions.

\subsection{Estimation of carcinogenic risk}
\label{ssec:estimation_carcinogenic_risk}
The estimated risk of developing lung cancer from inhalation of PAHs in the atmosphere can be computed based on an approach proposed by the World Health Organization (WHO) \mycitep{InternationalAgencyForResearchOnCancer2010}.
Benzo(a)pyrene (B[a]P) is used as a surrogate for PAHs, since it has been extensively studied in the literature and is considered carcinogenic to humans \mycitep{InternationalAgencyForResearchOnCancer2010}:
chronic exposure to $\SI{1}{\nano\gram\per\meter\cubed}$ B[a]P over a lifetime of 70 years is associated with a unit risk (UR) of $\mathrm{UR_{B[a]P}} = 8.7 \times 10^{-5}$ (in units $(\SI{}{\micro\gram\per\meter\cubed})^{-1}$), implying that 8.7 out of $100,000$ people may develop lung cancer from this exposure level.
However, estimating the carcinogenic risk of PAH mixtures is less well established. 
A common approach relies on assessing the potency of PAHs relative to B[a]P (e.g., \cite{Ramirez2011a}).
Similar to the calculation by the WHO, we thus estimate the lifetime carcinogenic risk as:
\begin{equation}
	\mathrm{Carcinogenic\ Risk} =
	\left( \sum_{i} \left[\mathrm{PAH}_{i}\right] \cdot \mathrm{TEF}_{\mathrm{PAH}_{i}} \right) \times \mathrm{UR}_{\mathrm{B[a]P}}
	\label{eqn:cr}
\end{equation}
where $\left[\mathrm{PAH}_{i}\right] = X_{\mathrm{PAH}_{i}}/X_{\ce{(C6H6)H+}}$ are individual (dimensionless) concentrations of PAHs normalized to benzene, and $\mathrm{TEF}_{\mathrm{PAH}_{i}}$ denotes the toxic equivalent factor (TEF) for a given PAH relative to B[a]P \citep{Ramirez2011a}.
By definition, B[a]P has a TEF of one, while other PAHs may have a fractional (lower potency compared to B[a]P) or larger (higher potency) value.
Further, we assume that the cancer risk of PAH mixtures is additive.
Using the TEFs proposed by \cite{Nisbet1992} (see Tab.~\ref{tab:TEF}), we obtain a relative carcinogenic risk of the sampled wildfire smoke in units $(\SI{}{\micro\gram\per\meter\cubed})^{-1}$, since we use PAH concentrations that are normalized to benzene.
Ultimately, we divide the resulting relative carcinogenic risk by its maximum value to obtain a normalized, and dimensionless carcinogenic risk. 
\begin{table*}
	\centering
 	\caption{Toxic equivalent factors (TEF) relative to Benzo(a)pyrene for relevant PAHs according to \cite{Nisbet1992}.}%
	\label{tab:TEF}
	\begin{tabularx}{0.82\textwidth}{@{}llll@{}}
		\hline
		Compound & Abbreviation & TEF \\ 
		\hline
		Naphthalene & NAP & 0.001\\
		Acenaphthylene & ACY & 0.001\\
		Acenaphthene & ACE & 0.001\\
		Fluorene & FLE & 0.001\\
		Phenanthrene/Anthracene & PH/AN & 0.001 \\
		Pyrene/Fluoranthene & PY/FLA & 0.001\\
		Benz(a)anthracene/Chrysene & B[a]A/CHR & 0.1\\
		Benzo(b)fluoranthene/Benzo(a)pyrene & B[b]F/B[a]P & 1\\
		\hline	\end{tabularx}

\end{table*}

\section{Results}
\label{sec:results}

\subsection{Gas-phase analysis}
\label{ssec:gas_phase results}
\begin{figure}
    \centering
    \includegraphics[width=0.8\linewidth]{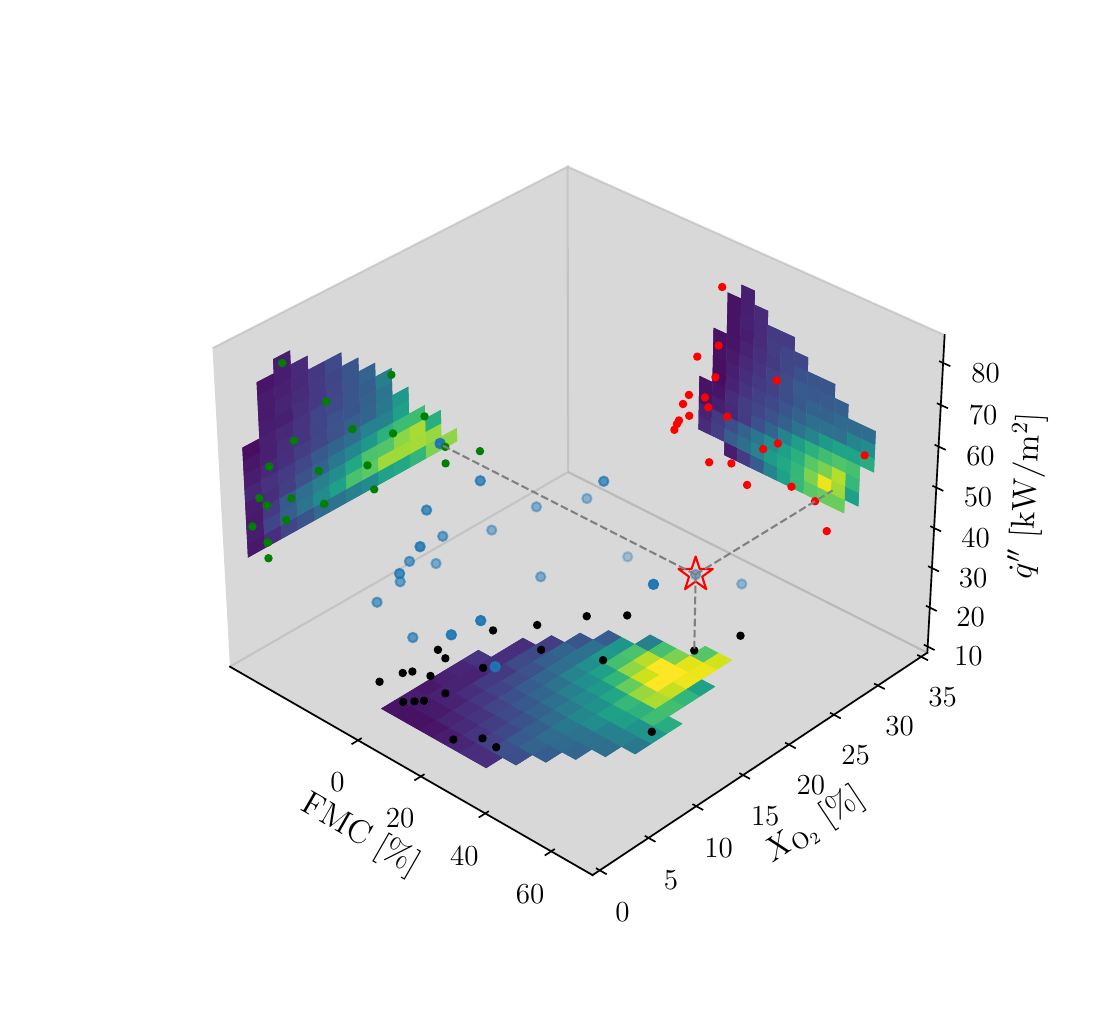}
    \caption{Selection of burn environment parameters FMC, $\dot{q}^{\prime\prime}$ and $X_{\ce{O2}}$ (blue markers).
	Each data point corresponds to one burn condition.
	The projected contour plots of phenanthrene/anthracene (PH/AN) emissions (normalized to benzene) are added for illustration. These are obtained through linear interpolation of burn-averaged emissions from all samples after projection onto 2D planes (green, red and black data points).
    The red star marks sample C (see Tab.~\ref{tab:sample_list} for reference).}
    \label{fig:sample_projection}
\end{figure}

\begin{figure*}
	\centering
	\includegraphics[width=.75\linewidth,trim={7cm 0 7cm .4cm},clip]{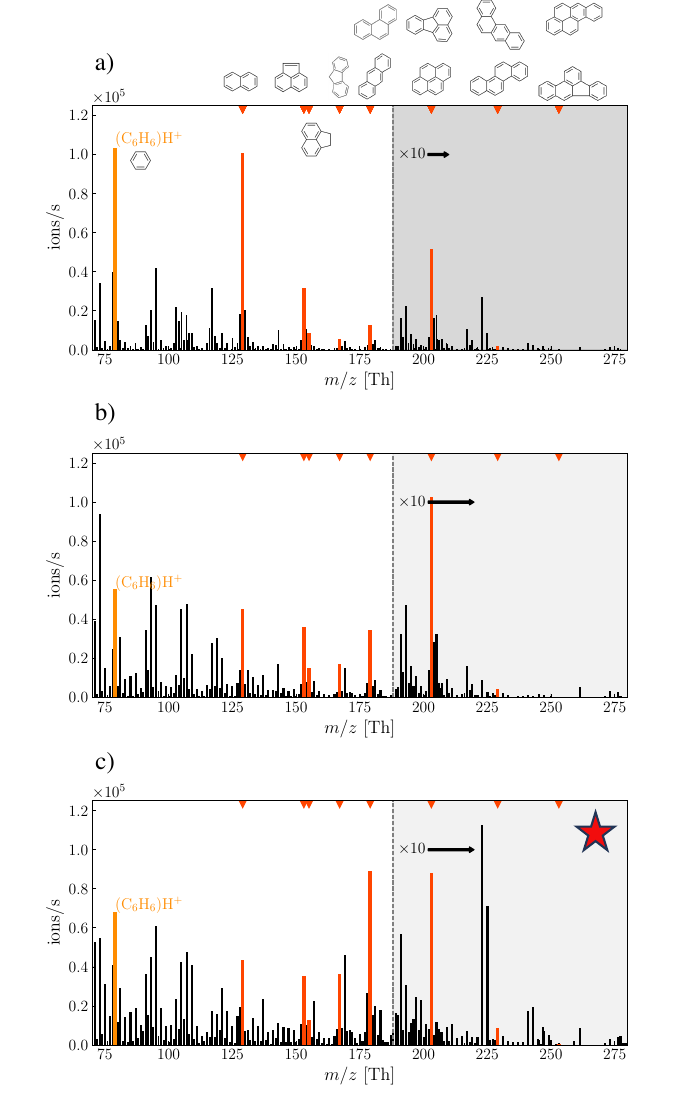}
	\caption{Burn averaged mass spectra of gas-phase VOCs for the samples A, B, and C, in panels (a) - (c), respectively.
		The PAHs considered in this work are highlighted in red at their detected ion mass given in $\SI{}{Th}$ (Thomson).
		Ion intensities of $m/z \geq \SI{180}{Th}$ (gray shaded region to the right of the vertical dashed line) are multiplied by a constant factor of 10 for readability.}
	\label{fig:mass_spectrum_E14}
\end{figure*}

All fuel samples considered in this work are visualized as blue circular markers in Fig.~\ref{fig:sample_projection} as a function of their local burn conditions.
For each burn, a burn-averaged mass spectrum is computed, from which average PAH concentrations are subsequently extracted.
The burn-averaged gas-phase mass spectra for three samples with distinctly different burn environments are shown in Fig.~\ref{fig:mass_spectrum_E14}. 
We refer to Tab.~\ref{tab:sample_list} for a specification of their burn conditions.
\begin{table}
	\centering
 \caption{Burn conditions for samples A, B, and C discussed in the text. Conditions for sample C are marked with a star in Fig.~\ref{fig:sample_projection}.}
 	\label{tab:sample_list}
	\begin{tabularx}{\textwidth}{@{}llll@{}}
		\hline
		Sample & FMC $[\%]$ & $X_{\mathrm{O}_2}\ [\%]$  & $\dot{q}^{\prime\prime}\ [\SI{}{\kilo\watt\per\metre\squared}]$ \\ 
		\hline
		A & 14.3 & 3.7 & 80.2 \\
		B & 33.6 & 14.4 & 63.9 \\
		C & 44.0 & 19.8 & 39.0 \\
		\hline	\end{tabularx}
\end{table}

A major part of the detected peaks can be associated with VOCs typically observed during biomass burns \citep{Koss2018}.
For the purpose of this work, however, we intentionally limit our gas-phase analysis to individual peaks between $m/z = \SI{79.054}{\thomson} - \SI{253.101}{\thomson}$ (highlighted bars in the mass spectrum of Fig.~\ref{fig:mass_spectrum_E14}) which we have tentatively identified as compounds listed in Tab.~\ref{tab:peak_list} \citep{Koss2018}.
\begin{table}
	\centering
 \caption{Ion peaks detected by PTR-ToF-MS and tentative identification of main VOC contributor.}%
	\label{tab:peak_list}
	\begin{tabularx}{\textwidth}{@{}llll@{}}
		\hline
		Peak $m/z$ & Ion formula & Main VOC contributor & Abbreviation\\ 
		\hline
		$\SI{79.054}{\thomson}$ & $\ce{(C6H6)H+}$ & Benzene \\
		$\SI{129.070}{\thomson}$ & $\ce{(C10H8)H+}$  & Naphthalene & NAP\\
		$\SI{153.070}{\thomson}$ & $\ce{(C12H8)H+}$ & Acenaphthylene & ACY\\
		$\SI{155.086}{\thomson}$ & $\ce{(C12H10)H+}$ & Acenaphthene & ACE\\
		$\SI{167.086}{\thomson}$ & $\ce{(C13H10)H+}$ & Fluorene & FLE\\
		$\SI{179.086}{\thomson}$ & $\ce{(C14H10)H+}$ & Phenanthrene/Anthracene & PH/AN \\
		$\SI{203.086}{\thomson}$ & $\ce{(C16H10)H+}$ & Pyrene/Fluoranthene & PY/FLA\\
		$\SI{229.101}{\thomson}$ & $\ce{(C18H12)H+}$ & Benz(a)anthracene/Chrysene & B[a]A/CHR\\
		\multirow{2}*{$\SI{253.101}{\thomson}$} & \multirow{2}*{$\ce{(C20H12)H+}$} & Benzo(b)fluoranthene/ & \multirow{2}*{B[b]F/B[a]P} \\
            & &  Benzo(a)pyrene & \\
		\hline	
    \end{tabularx}
\end{table}
Except for benzene, these compounds are considered as Priority PAHs by the U.S. EPA \citep{Yan2004}.
Note that the Vocus does not differentiate between isomers at a given peak, thus we assume that all isomers contribute to the total ion count at their corresponding $m/z$.
We will henceforth refer to the detected ions by their assumed VOC contributor \citep{Koss2018}.

A distinct pattern can be observed across most burn-averaged spectra when comparing concentrations of individual PAHs for a given burn environment: 
emission profiles are dominated by NAP and PH/AN (among the considered PAHs), reaching averaged intensities that are of the same order of magnitude as benzene under most conditions, as illustrated for samples B and C (Fig.~\ref{fig:mass_spectrum_E14}b and c).
For the remaining PAHs, intensities decrease with increasing $m/z$ values, with ACY, ACE, FLE showing consistently lower concentrations relative to benzene than the more stable NAP, PH/AN.
For $m/z = \SI{229.101}{\thomson}$ and above, ion intensities in the gas-phase are very low and virtually indistinguishable from background noise.
In turn, sample A deviates from this pattern, as NAP is found to be most dominant compared to all other PAHs.
Note, however, that ion counts for a single PAH cannot be compared directly across different burn conditions (represented by the spectra in Fig.~\ref{fig:mass_spectrum_E14}), since the dilution with ambient air is not kept constant for each burn.
Therefore, we normalize PAH concentrations with benzene concentrations of the same burn (similar to the approach by \cite{ODell2020}), before comparing burn-to-burn variations of PAH emissions.
This procedure facilitates an unbiased evaluation of a given burn environment in terms of the generated PAH emissions, and can further aid an assessment of potential health risks.
We compared benzene normalization to HCN normalization which did not change our conclusions.

\begin{figure}
	\centering
	\includegraphics[width=0.8\linewidth]{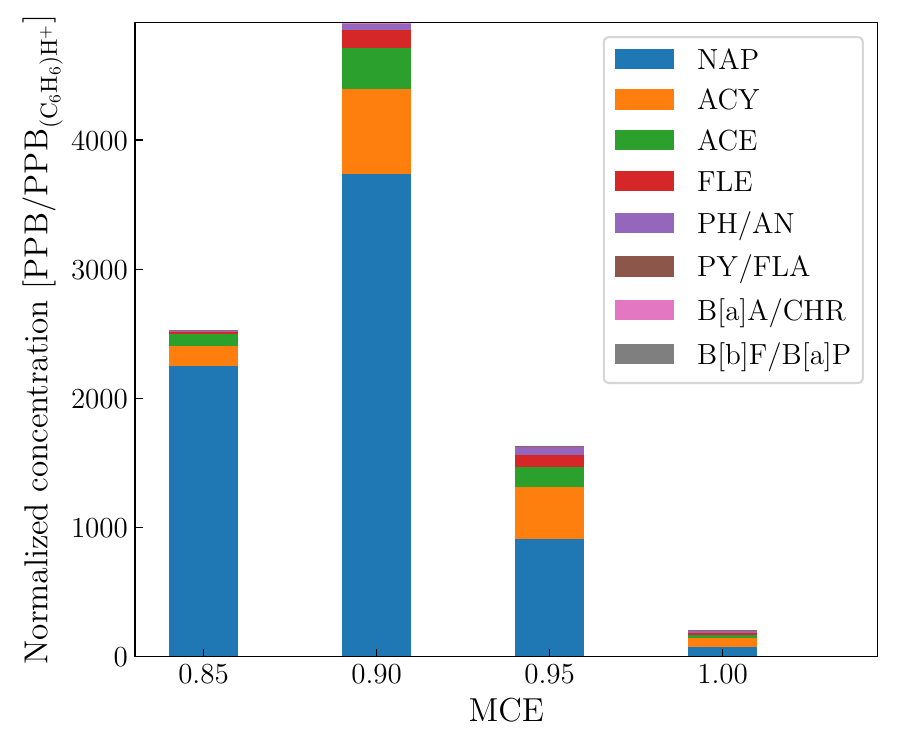}
	\caption{Normalized concentrations to $\ce{(C6H6)H+}$ as a function of the modified combustion efficiency, and averaged over all samples.}
	\label{fig:MCE_bar_chart}
\end{figure}

In Fig.~\ref{fig:MCE_bar_chart}, we investigate whether normalized PAH concentrations correlate with the instantaneous MCE as an overall characterization of the combustion process.
Each bar is obtained by averaging the emissions from all samples that fall into a given bin (bin width is $\pm 0.025$ of the indicated value).
Highest PAH concentrations are found during periods of low MCE values (\emph{i.e.,} $\mathrm{MCE} \lesssim 0.9$) at the beginning of each burn, 
suggesting that a predominantly smoldering combustion regime enhances PAH formation.

In general, precursors for PAH formation are formed during pyrolysis, in reactions that occur within a temperature range of $T \approx 600 - \SI{900}{\kelvin}$ \citep{Anca-Couce2016}.
Despite the much higher burn environment temperatures considered in this study ($T > \SI{1476}{\kelvin}$), the pyrolysis process may be characterized by the thermally thick regime~\citep{Anca-Couce2016} due to comparably large sample diameters and large Biot numbers ($\mathrm{Bi} = hd/(2k) \approx 25 \gg 1$). 
Therefore, the reaction rate of pyrolysis is controlled by internal heat transfer, which we can influence by controlling the heat load onto a sample.

Increasing the pyrolysis temperature by increasing the heat load onto the sample enhances aromatization of pyrolyzates \citep{Sekimoto2018}, contributing to the generation of light cyclic hydrocarbons such as benzene and NAP. 
The thermally thick regime also promotes the formation of tars \citep{Bennadji2013} containing important precursors, such as phenyl radicals \citep{Liu2019a}, from which higher-weight PAHs can be formed.
Finally, the subsequent combustion process of the pyrolizate critically impacts the fate of the initial products depending on the chemical reaction pathways \citep{Wang2011a} controlling the growth of hydrocarbons.
Therefore, we anticipate PAH emissions to be primarily governed by (i) the heat load onto the sample, and (ii) the combustion temperature indirectly controlled by the sample's FMC.

\begin{figure}
    \centering
    \includegraphics[width=\linewidth]{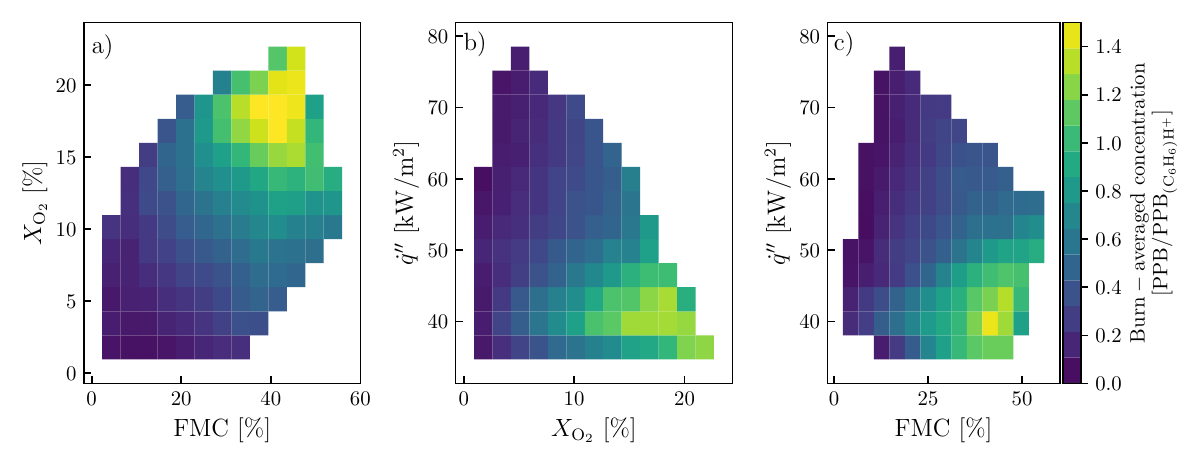}
    \caption{Burn-averaged emission of phenanthrene/anthracene (PH/AN) normalized to benzene as a function of the burn environment.
    Panel a) shows the projected emission map as a function of $X_{\ce{O2}}$ and FMC, panel b) as function of $\dot{q}^{\prime\prime}$ and $X_{\ce{O2}}$, and panel c) as function of $\dot{q}^{\prime\prime}$ and FMC.
    }
    \label{fig:heat_map_phenanthrene}
\end{figure}

To reveal the effect of these parameters on resulting PAH concentrations we compute ``emission maps'' as follows:
for each PAH, we compute normalized burn-averaged concentrations as a function of the corresponding burn conditions.
We then project the concentrations onto a two-dimensional parameter sub-space given by $\dot{q}^{\prime\prime}$ vs. $X_{\mathrm{O}_2}$, $\dot{q}^{\prime\prime}$ vs. FMC, and FMC vs. $X_{\mathrm{O}_2}$, corresponding to the projected green, red and black markers in Fig.~\ref{fig:sample_projection}, respectively, to facilitate our analysis.
Ultimately, the emission maps are obtained by interpolating the projected data.
We focus on PH/AN in Fig.~\ref{fig:heat_map_phenanthrene} for clarity, although very similar patterns can be observed for ACY to PY/FLA (see Fig.~\ref{fig:heat_maps_acenaphthylene} to \ref{fig:heat_maps_fluoranthene_pyrene}).
From a health perspective, PH exposure deserves particular attention as it alters the immunologic response \citep{Liu2013}.

From Fig.~\ref{fig:heat_map_phenanthrene}, we first observe that the burn-averaged emissions of PH/AN span a range of 76:1 within the parameter space considered in our work.
The effect of FMC, heat flux, and oxygen deprivation on PAH emissions therefore appears to be significantly stronger than the more widely studied effect of fuel type: 
\cite{Jenkins1996} report a factor of 2 across 4 species of woody biomass studied, while \cite{Fine2001} observed a variation by a factor of 8 across 6 species.

Inspecting the burn-averaged emissions as a function of the heat flux $\dot{q}^{\prime\prime}$ onto the sample and FMC (Fig.~\ref{fig:heat_map_phenanthrene}c),
two distinct regions become apparent:
$\mathrm{FMC} \geq 30\%$ tends to enhance PH/AN (relative to benzene) if the heat flux remains sufficiently low ($\dot{q}^{\prime\prime} \lesssim \SI{55}{\kilo\watt\per\meter\squared}$), suggesting competing effects between both parameters.
In this regime, a high FMC leads to lower combustion temperatures due to evaporation, attenuating PAH fragmentation \citep{Wang2011a}, and thus enhancing the generation of higher-weight PAHs from their precursors.
Conversely, an augmented heat flux to the sample ($\dot{q}^{\prime\prime} \geq \SI{55}{\kilo\watt\per\meter\squared}$) resulting in faster evaporation and stronger fragmentation substantially reduces the concentrations of higher-weight PAHs relative to benzene.
This suggests that burn environments with a heat flux greater than $\dot{q}^{\prime\prime} \gtrsim \SI{55}{\kilo\watt\per\meter\squared}$ are beneficial for attenuating normalized emissions of PH/AN by 77\% compared to burn environments characterized by lower heat fluxes.
A similar trend is also observed for the remaining PAHs, albeit with a lower emission reduction (see \ref{sec:heat_maps_appendix}).

At FMC below 20\%, the impact of evaporation becomes weaker within the explored parameter range.
Consequently, high heat loads lead to a reduced formation of larger PAHs from their pyrolysis precursors.
NAP (see Fig.~\ref{fig:heat_maps_naphthalene}) deviates from the trends described above, exhibiting the least variability across the explored parameter range.
Its formation is likely facilitated by the abundance of benzene originating from pyrolysis.

\begin{figure}
			\centering
			\includegraphics[width=.8\linewidth]{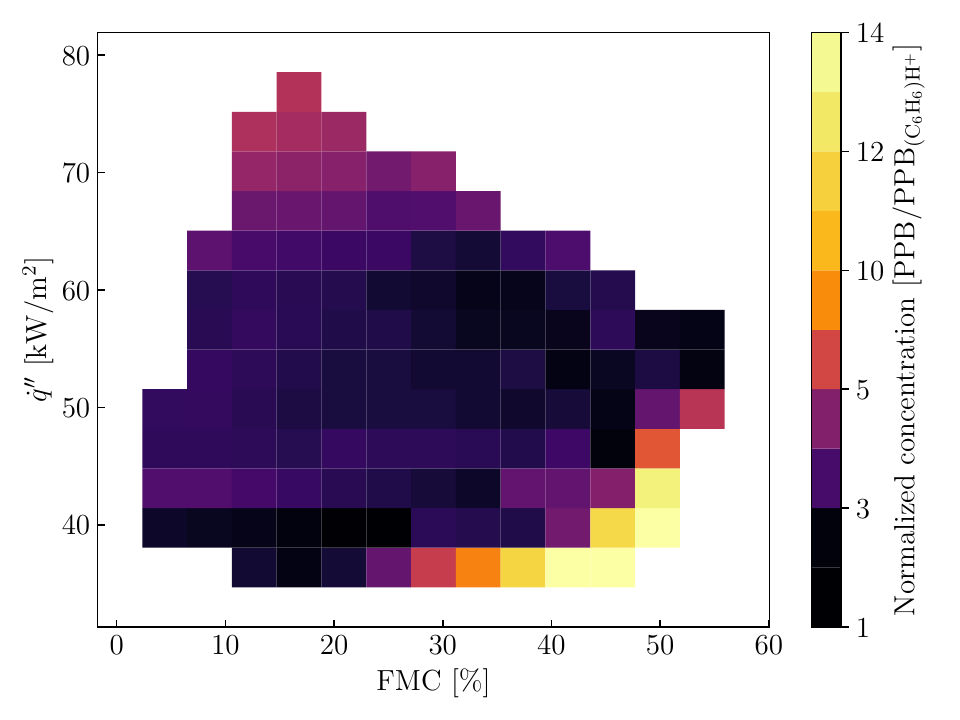}
			\caption{Burn-averaged concentration of acetylene ($\ce{C2H2}$) normalized to benzene. 
			The colorbar is given in log-scale.}
			\label{fig:heat_map_acetylene}
\end{figure}

Apart from the effect of FMC, other mechanisms could be invoked to explain the decrease of PAH emissions with increasing heat load.
Figure~\ref{fig:heat_map_acetylene} shows the concentration of acetylene ($\ce{C2H2}$, measured using TILDAS, normalized to benzene), exhibiting a distinct peak at high FMC and low heat loads, which coincides with the high-emission region of ACY and higher molar weight PAHs, i.e., $\mathrm{FMC} \gtrsim 30\%$ and $\dot{q}^{\prime\prime} \lesssim \SI{55}{\kilo\watt\per\metre\squared}$.
Acetylene is a pyrolysis product \citep{Martin2022} and may form in significant quantities during biomass pyrolysis \citep{Zhang2007a} at temperatures above $\SI{970}{\kelvin}$.
Since the presence of acetylene may facilitate the formation and growth of PAHs \citep{Frenklach1985}, it can also explain the observed peak PAH emissions in this region.
Moreover, normalized $\ce{C2H2}$ concentrations reach a minimum at intermediate heat loads, but increase again at $\dot{q}^{\prime\prime} \gtrsim \SI{70}{\kilo\watt\per\metre\squared}$.
This region of increased $\ce{C2H2}$ concentrations could explain observed particle-phase emission patterns discussed in the following section.
	
\subsection{Particle-phase analysis}
\begin{figure}
    \centering
    \includegraphics[width=0.7\linewidth]{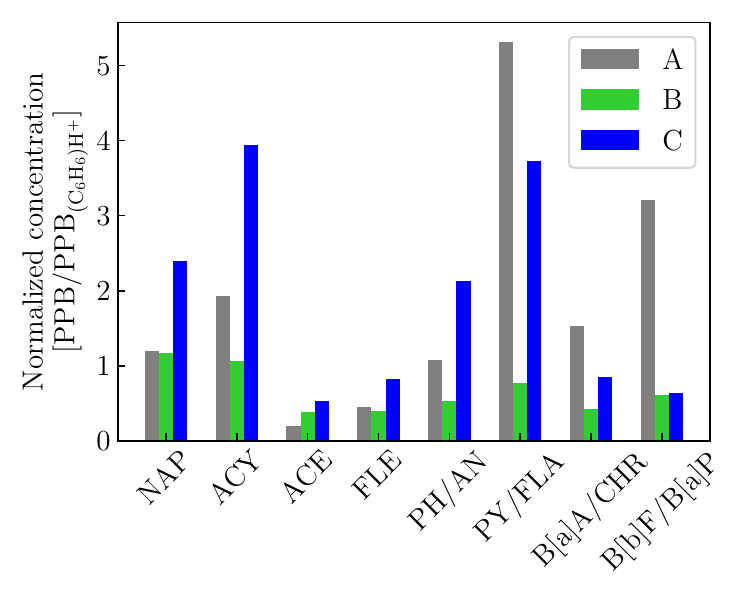}
    \caption{Particle-phase PAH concentrations normalized to benzene (in the gas-phase) for three individual burns labeled as A, B, and C, with their corresponding burn parameters summarized in Tab.~\ref{tab:sample_list}.}
    \label{fig:particle_emissions}
\end{figure}

A significant fraction of the PAHs condensates onto the particle-phase, which must therefore be studied alongside the gas-phase to provide a comprehensive assessment of PAH emissions in a given burn environment. 
This is particularly true for higher molecular weight PAHs \citep{Lima2005}.
Thus, for samples A, B, and C, corresponding to low, intermediate, and high FMC (see Tab.~\ref{tab:sample_list}), we investigated particle-phase compositions with HR-ToF-SP-AMS.
Concentrations normalized to gas-phase benzene are shown in Fig.~\ref{fig:particle_emissions}.
Two major conclusions can be drawn:
(i) normalized signal intensities of PY/FLA and higher molar weight PAHs are at least one order of magnitude larger compared to their gas-phase counterparts (for sample C: 3.68\,PPB/PPB$_{(\mathrm{C_6H_6)H^{+}}}$ in the particle-phase, Fig.~\ref{fig:particle_emissions}, versus 0.16\,PPB/PPB$_{(\mathrm{C_6H_6)H^{+}}}$ in the gas phase, Fig.~\ref{fig:heat_maps_fluoranthene_pyrene}), indicating that these compounds are formed during the combustion process, but tend to condensate onto the particle phase unlike low molar weight PAHs;
(ii) concentrations of particle-phase PAHs are noticeably low at intermediate fuel moisture (sample B), but increase towards lower FMC (sample A), as well as higher FMC (sample C).

\section{Discussion}
\label{sec:discussion}
\subsection{Implications for selecting an optimal burn environment based on carcinogenic risk}
The combined gas-phase and particle-phase PAH emission data allows us to evaluate  burn environments in terms of overall PAH emissions and expected health impacts from smoke inhalation.
\begin{figure}
    \centering
    \includegraphics[width=.7\textwidth,trim={6.5cm 8.5cm 6.5cm 0},clip]{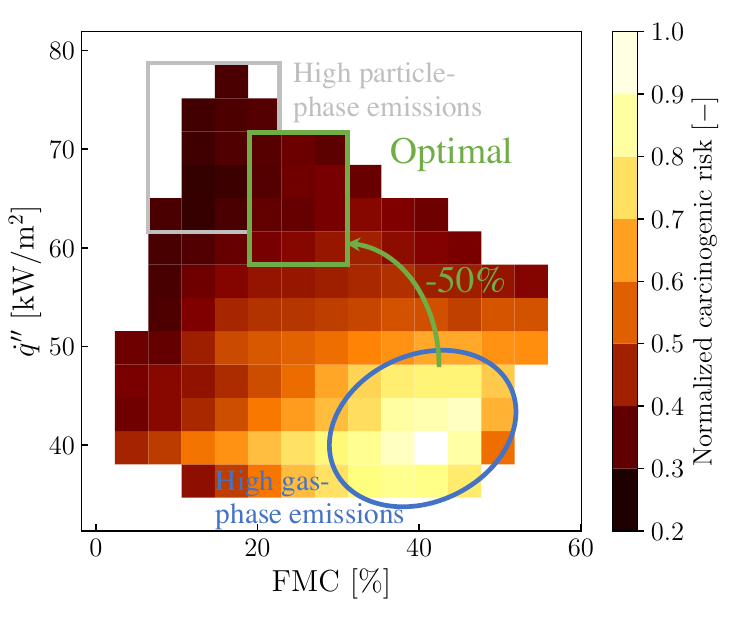}
    \caption{Assessment of burn environments based on the relative carcinogenic risk.
	Burn environments in the blue ellipse should be avoided due to high gas-phase emissions, while the region within the grey rectangle should be avoided due to high particle-phase emissions.
	The recommended burn environment is highlighted by a green rectangle as it features both low gas and particle-phase PAH emissions.
	The normalized carcinogenic risk reduces by more than 50\% (associated with gas-phase PAHs) if burns are performed in the optimal burn window.}
    \label{fig:carcinogenic_risk_map}
\end{figure}
As detailed in Sec.~\ref{ssec:estimation_carcinogenic_risk}, we have converted the gas-phase PAH emissions from all samples into a normalized carcinogenic risk (see contour plot in Fig.~\ref{fig:carcinogenic_risk_map}) to aid the selection of an optimal burn environment.
The region at $\mathrm{FMC} \gtrsim 30\%$ and $\dot{q}^{\prime\prime} \leq \SI{55}{\kilo\watt\per\meter\squared}$ shows the overall highest carcinogenic risk within the explored parameter range (blue ellipse in Fig.~\ref{fig:carcinogenic_risk_map}), associated with peak emissions of the investigated gas-phase PAHs.
Since burn environments in this region also promote high particle-phase PAH concentrations (inferred from sample C, see Fig.~\ref{fig:particle_emissions}), they are expected to be most concerning from a health perspective.
In turn, while the formation of most gas-phase PAHs is attenuated at low FMC and high heat flux (grey rectangle in Fig.~\ref{fig:carcinogenic_risk_map}), this regime should be avoided due to its elevated particle-phase emissions (inferred from sample A).
This leaves an optimal parameter range for the burn environment found within $20\% \leq \mathrm{FMC} \leq 30\%$, $\SI{60}{\kilo\watt} \leq \dot{q}^{\prime\prime} \leq \SI{70}{\kilo\watt\per\meter\squared}$ (see green rectangle in Fig.~\ref{fig:carcinogenic_risk_map}), and an oxygen concentration between $5$ and $15\%$ for the considered fuel species.
Burn environments within this parameter range simultaneously achieve low gas and particle-phase (inferred from sample B) emissions, which could contribute to reducing the carcinogenic risk associated with inhalation of PAHs from prescribed fires.
Other factors not considered in our idealized setup could however alter the above parameter range, the resulting carcinogenic risk, or both.
Future studies should therefore investigate the parametric sensitivities of additional factors affecting PAH formation, with the present work being a first step in that direction.

Despite these limitations, the present approach can aid a more comprehensive evaluation of benefits of a planned prescribed fire activity by including expected PAH emissions as an additional criterion, alongside variations in the local burn environment.
Similarly, if burn conditions are expected to be outside the optimal parameter range, additional pre-burn fuel manipulation could be considered to perform the prescribed burn at more desirable conditions.
Our insights could further influence the selection of the most appropriate treatment (\emph{e.g.,} broadcast burn, pile burn, or pit burners), since different burn types offer a varying degree of control over the burn environment.

Although larger broadcast burns offer the least amount of direct control over the resulting burn environment, burn parameters may still be adjusted through indirect measures:
weather forecasts can provide sufficiently accurate FMC data, allowing to plan the day-of-burn when fuel moisture is expected to be within the optimal parameter range.
This is already today an established procedure for planning prescribed fires and mitigating fire escape \citep{SmokeManagement}.
The heat flux onto the fuel could be affected more indirectly by altering the available fuel density through appropriate pre-burn manipulation attempting to change the fuel arrangement or size with mechanical tools \citep{Pique2018}.
These measures may also be beneficial for ensuring a sufficient oxygen flux into the burn environment.

If broadcast burns are deemed unfeasible or unsafe, pile burns could be considered as an alternative as they can increase the degree of control over the burn environment, as demonstrated by \emph{e.g.,} \cite{Aurell2017}.
Prior to pile burning, fuel is often processed by mechanical or power tools to generate the desired fuel arrangement and size, and thereby more directly the resulting heat flux during the burn.

Lastly, mobile applications such as pit burners \citep{Miller2007}, incinerators, or air-curtain burners \citep{Lee2017} offer the highest degree of control over the burn environment as they closely resemble the laboratory-scale burner presented in this work.
This type of burn would be most attractive in the vicinity of densely populated areas, where excessive air pollution should be avoided to the maximum extent possible in those areas.

\subsection{Limitations of the present work}
Although the above recommendations offer considerable reductions of expected PAH emissions, they should not be considered in an isolated manner or as sole criterion to evaluate the benefits of planned prescribed fire activities.
We have intentionally limited our study to a subset of carcinogenic air pollutants to observe the effect of the burn environment on expected concentration levels.
Other pollutants with negative health outcomes have not been considered in the present work, and optimal burn conditions for minimal PAH emissions do not necessarily coincide with optimal conditions for other pollutants.
Further studies are needed to investigate the emission behavior of other pollutants as a function of the underlying burn environment.

Apart from the inherent health risk, other social factors not discussed in this work should be included in the overall evaluation of prescribed burns.
We reiterate that prescribed fire activities contribute to air pollution and may expose some parts of the population to unhealthy air quality more often than others due to socio-economic factors \citep{Afrin2021}.
However, prescribed burns are commonly planned several days or weeks in advance and timely alerts can be issued to affected communities, alongside recommendations to stay indoors, wear masks, or pursue active measures to improve indoor air quality \citep{Liang2021}.

\section{Conclusions}
Prescribed fire activity constitutes an important cornerstone of forest management and wildfire risk mitigation.
In light of an increased use of controlled burns, populations are more likely to be exposed to unhealthy levels of air pollution, highlighting the need to improve our understanding of potential health risks arising from toxic and carcinogenic pollutants found in the smoke of biomass combustion.
Thus, we have performed laboratory-scale burns of Eastern White Pine under relevant burn conditions to demonstrate that an optimal burn environment can reduce pollutant emissions and thereby limit negative health outcomes.
We have focused our analysis on polycyclic aromatic hydrocarbons (PAHs), which have significant detrimental health impacts.
Further, there remains significant variability in their emission levels from biomass combustion reported in  the literature.

Our well-controlled laboratory-scale combustor setup allows for independent variation of the heat load onto the sample, oxygen concentration in the burn environment, and sample fuel moisture content.
Gas-phase burn emissions were sampled via a time-resolved proton-transfer time-of-flight mass spectrometer (PTR-ToF-MS), while particle-phase emissions were characterized by a High-Resolution Time-of-Flight Soot-Particle Aerosol Mass Spectrometer (HR-ToF-SP-AMS). 

Our results suggest that the sample fuel moisture content, the heat load onto the sample, and the oxygen concentration in the burn environment have a very significant impact on the emission levels of PAHs, with emissions of phenanthrene varying over a span of 76:1 in our experiments. 
This observed variability is an order of magnitude larger than the variability originating from the species of biomass in \cite{Jenkins1996} or \cite{Fine2001}.
Burns conducted within the optimal burn parameter range have the potential to reduce normalized emissions by up to 77\%, corresponding to a reduction of the normalized carcinogenic risk by more than 50\%.
We then discuss viable strategies to apply our findings in real-world scenarios within the limitations of other factors, which we have not varied in this work.
Future work will focus on extending the present methodology to other types of fuels.

\section*{Acknowledgements}
Financial support through the Stanford Sustainability Accelerator, the Google Research Scholar program, and the Moore Foundation is gratefully acknowledged.
MFK’s work at SLAC is supported by the U.S. Department of Energy, Office of Science, Basic Energy Sciences, Scientific User Facilities Division, under Contract No. DE-AC02-76SF00515.
The authors thank Andrew Lambe, Megan Claflin, and Brian Lerner (Aerodyne Research) for their support during the experiment and helpful discussions.

 \FloatBarrier
 \bibliographystyle{elsarticle-harv}
\bibliography{aerodyne_CNF}

\newpage

\appendix
\setcounter{figure}{0}
\section{Peak fit}
\label{sec:peak_fit_appendix}
\begin{figure}
			\centering
			\includegraphics[width=\linewidth]{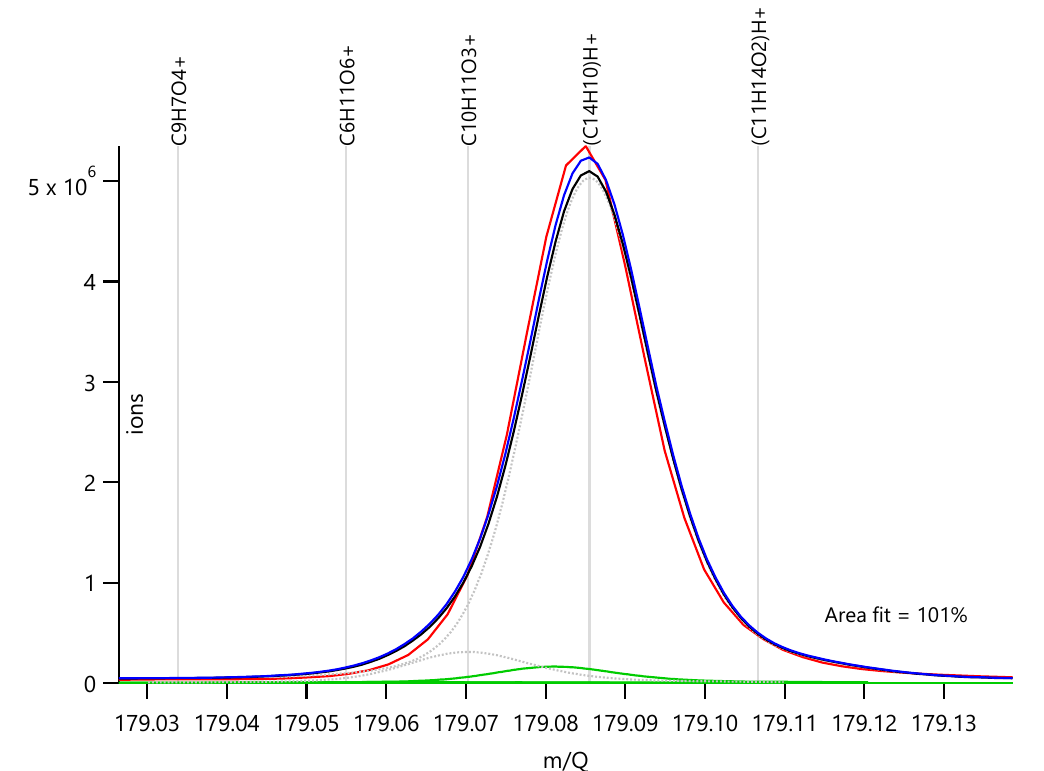}
			\caption{Example of the fully constrained peak fit shown for $\ce{(C14H10)H+}$ (phenanthrene/anthracene (PH/AN)).}
			\label{fig:peak_fit_igor}
		\end{figure}

Analysis of the raw mass spectrometer data was performed in Tofware v3.3.0 with IgorPro v9.0.2.4.
An example of the fully constrained peak fit is shown in Fig.~\ref{fig:peak_fit_igor} for $\ce{(C14H10)H+}$ which we have tentatively identified as phenanthrene/anthracene (PH/AN).

\setcounter{figure}{0}
\section{Emission profiles for investigated PAHs in the gas phase}
\label{sec:heat_maps_appendix}
\begin{figure*}
			\centering
			\begin{subfigure}[c]{0.05\linewidth}
				\includegraphics[scale=.035]{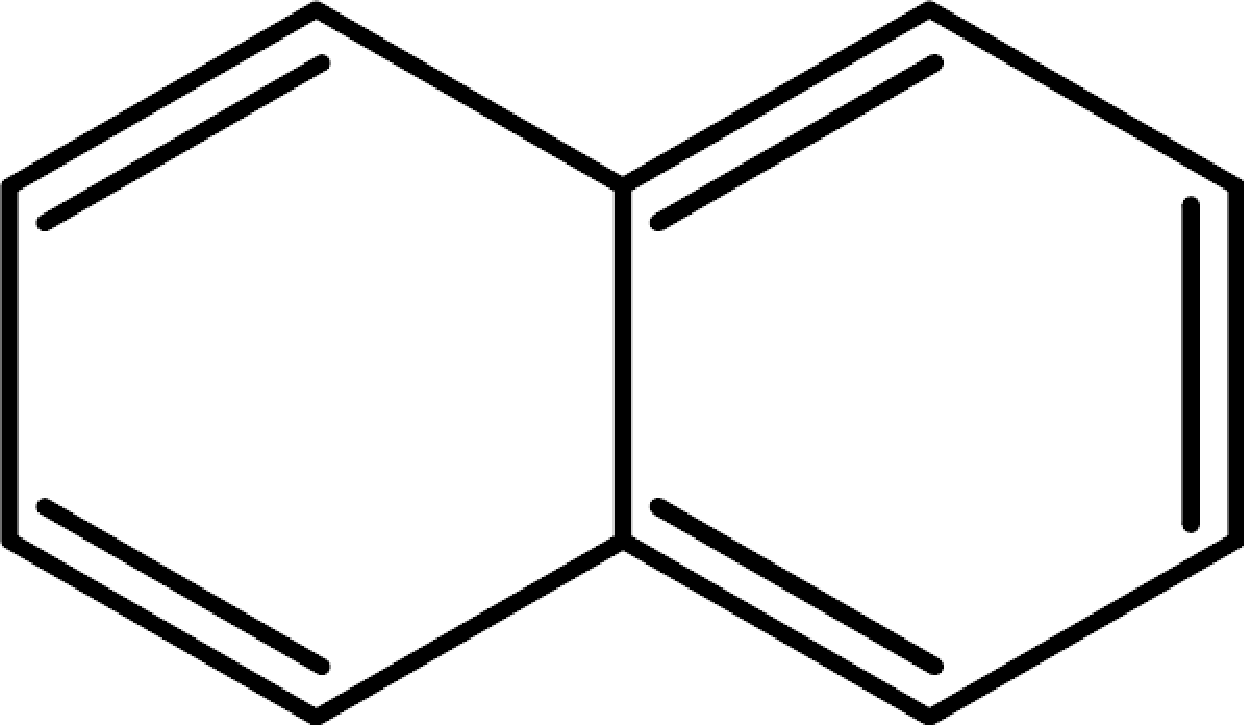}
			\end{subfigure}
			\begin{subfigure}[c]{0.94\linewidth}
				\includegraphics[width=\textwidth]{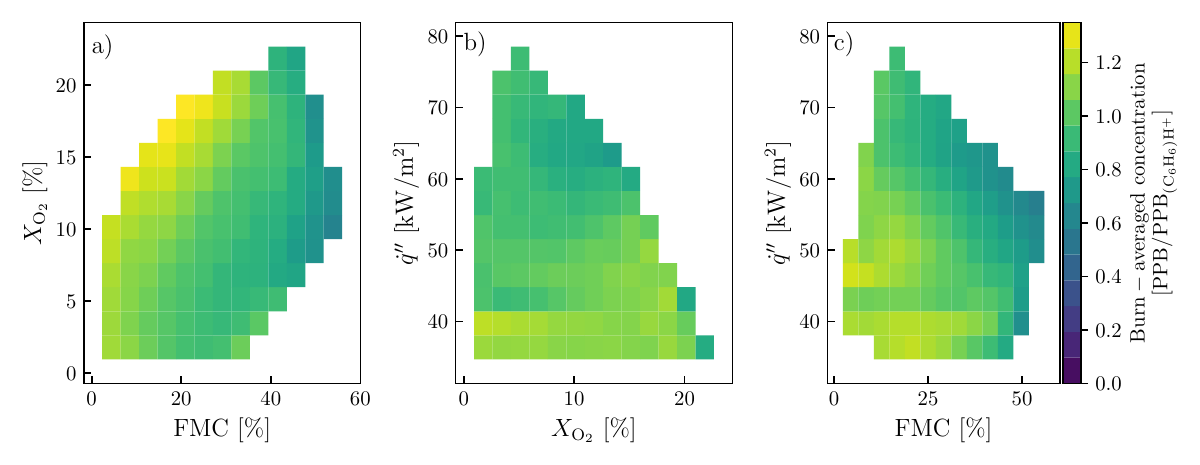}		
			\end{subfigure}
			\caption{Averaged emission of naphthalene (NAP) normalized to benzene as a function of the burn environment.
				Gas-phase emissions are reduced by 32.4\% if burn conditions are chosen in the optimal parameter range compared to conditions for sample C.}
			\label{fig:heat_maps_naphthalene}
		\end{figure*}

\begin{figure*}
			\centering
			\begin{subfigure}[c]{0.05\linewidth}
				\includegraphics[scale=.22]{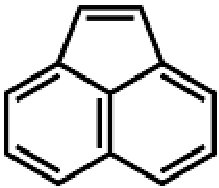}
			\end{subfigure}
			\begin{subfigure}[c]{0.94\textwidth}
				\includegraphics[width=\textwidth]{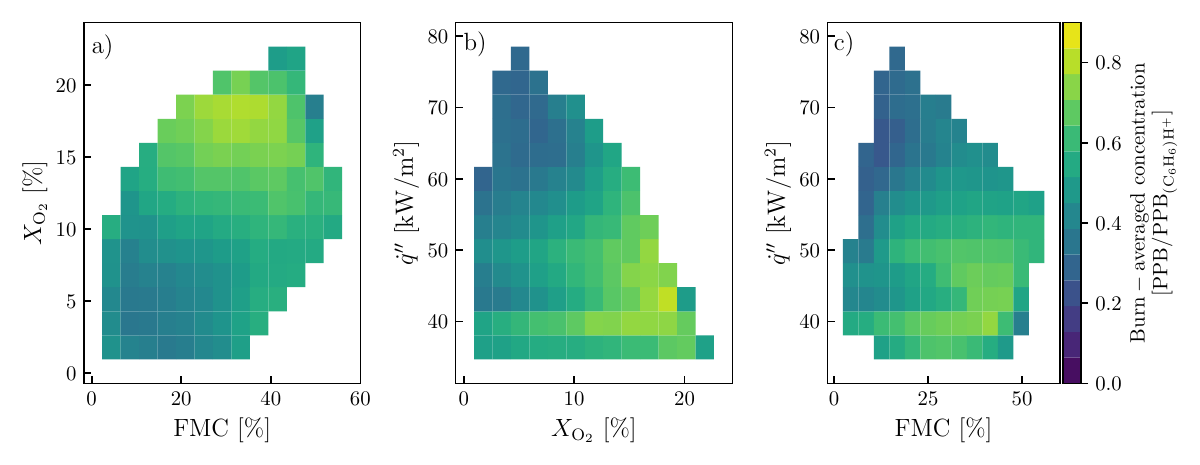}
			\end{subfigure}
			\caption{Averaged emission of acenaphthylene (ACY) normalized to benzene as a function of the burn environment.
				Gas-phase emissions are reduced by 40.9\% if burn conditions are chosen in the optimal parameter range compared to conditions for sample C.}
			\label{fig:heat_maps_acenaphthylene}
		\end{figure*}

\begin{figure*}
			\centering
			\begin{subfigure}[c]{0.05\linewidth}
				\includegraphics[scale=.14]{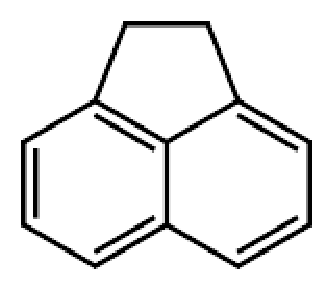}
			\end{subfigure}
			\begin{subfigure}[c]{0.94\textwidth}
				\includegraphics[width=\textwidth]{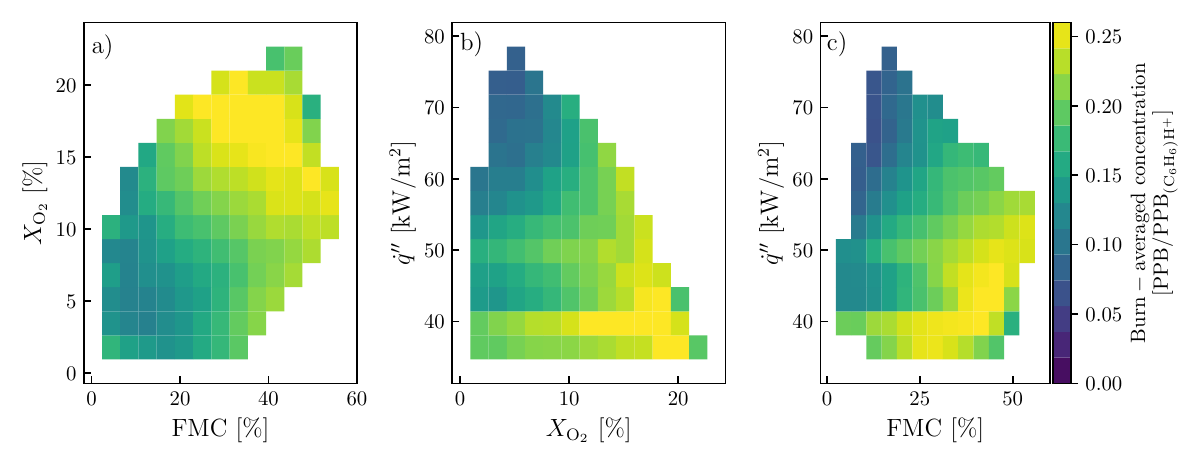}
			\end{subfigure}
			\caption{Averaged emission of acenapthene (ACE) normalized to benzene as a function of the burn environment.
				Gas-phase emissions are reduced by 31.9\% if burn conditions are chosen in the optimal parameter range compared to conditions for sample C.}
			\label{fig:heat_maps_acenaphthene}
		\end{figure*}

\begin{figure*}
			\centering
			\begin{subfigure}[c]{0.05\linewidth}
				\includegraphics[scale=.08]{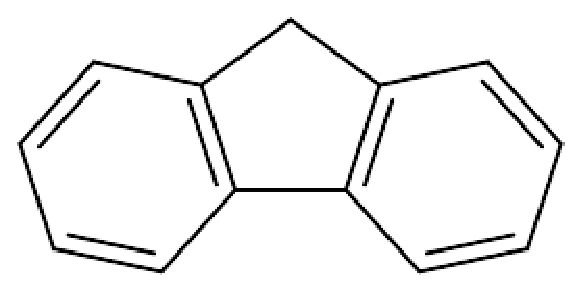}
			\end{subfigure}
			\begin{subfigure}[c]{0.94\textwidth}
				\includegraphics[width=\textwidth]{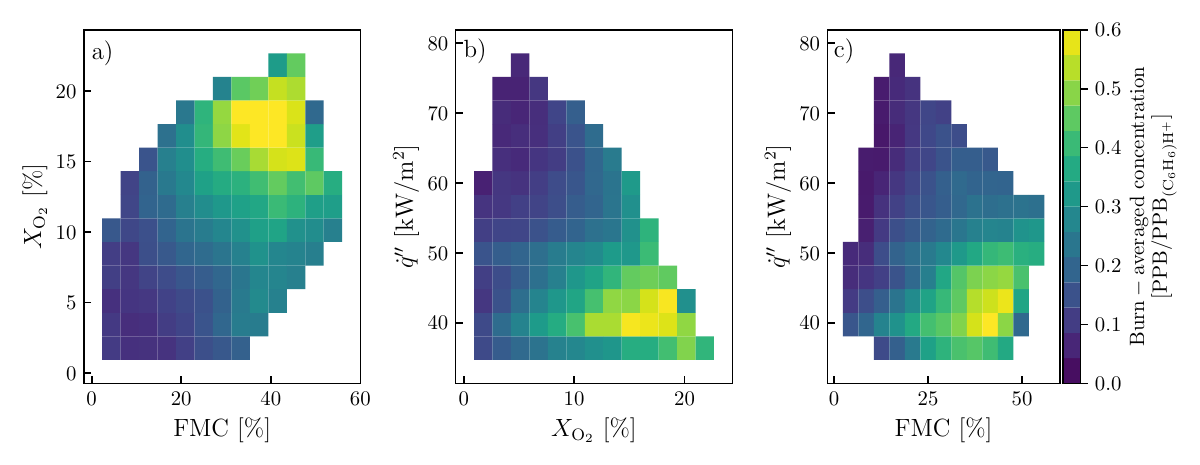}
			\end{subfigure}
			\caption{Averaged emission of fluorene (FLE) normalized to benzene as a function of the burn environment.
				Gas-phase emissions are reduced by 73.2\% if burn conditions are chosen in the optimal parameter range compared to conditions for sample C.}
			\label{fig:heat_maps_fluorene}
		\end{figure*}

\begin{figure*}
			\centering
			\begin{subfigure}[c]{0.05\linewidth}
				\includegraphics[scale=.13]{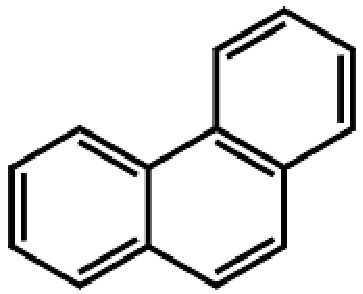}
				\includegraphics[scale=.028]{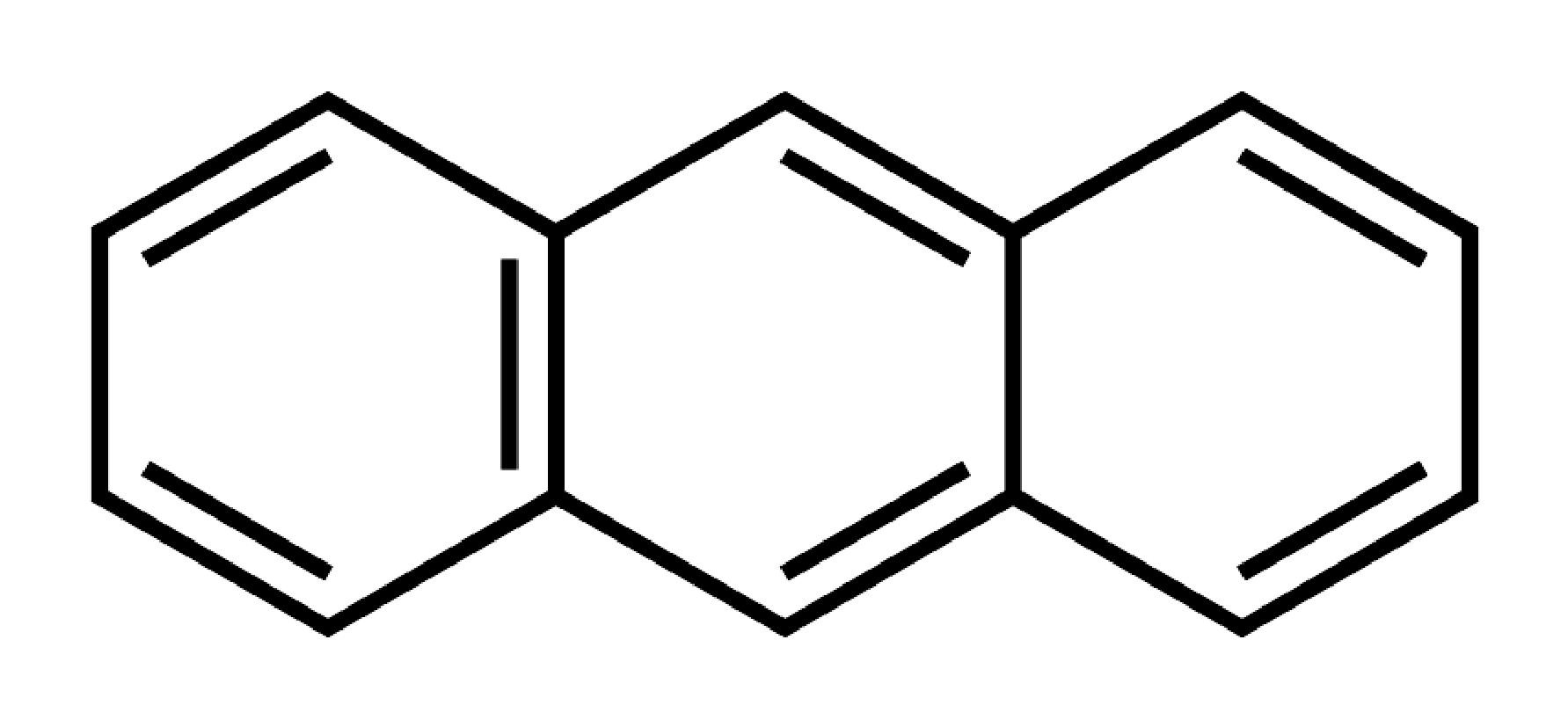}
			\end{subfigure}
			\begin{subfigure}[c]{0.94\textwidth}
				\includegraphics[width=\textwidth]{img/heat_map_phenanthrene_protonated_norm_ion_benzene_combined.pdf}
			\end{subfigure}
			\caption{Averaged emission of phenanthrene/anthracene (PH/AN) normalized to benzene as a function of the burn environment.
				Gas-phase emissions are reduced by 77.5\% if burn conditions are chosen in the optimal parameter range compared to conditions for sample C.}
			\label{fig:heat_maps_phenanthrene_anthracene_SI}
		\end{figure*}

\begin{figure*}
			\centering
			\begin{subfigure}[c]{0.05\linewidth}
				\includegraphics[scale=.043]{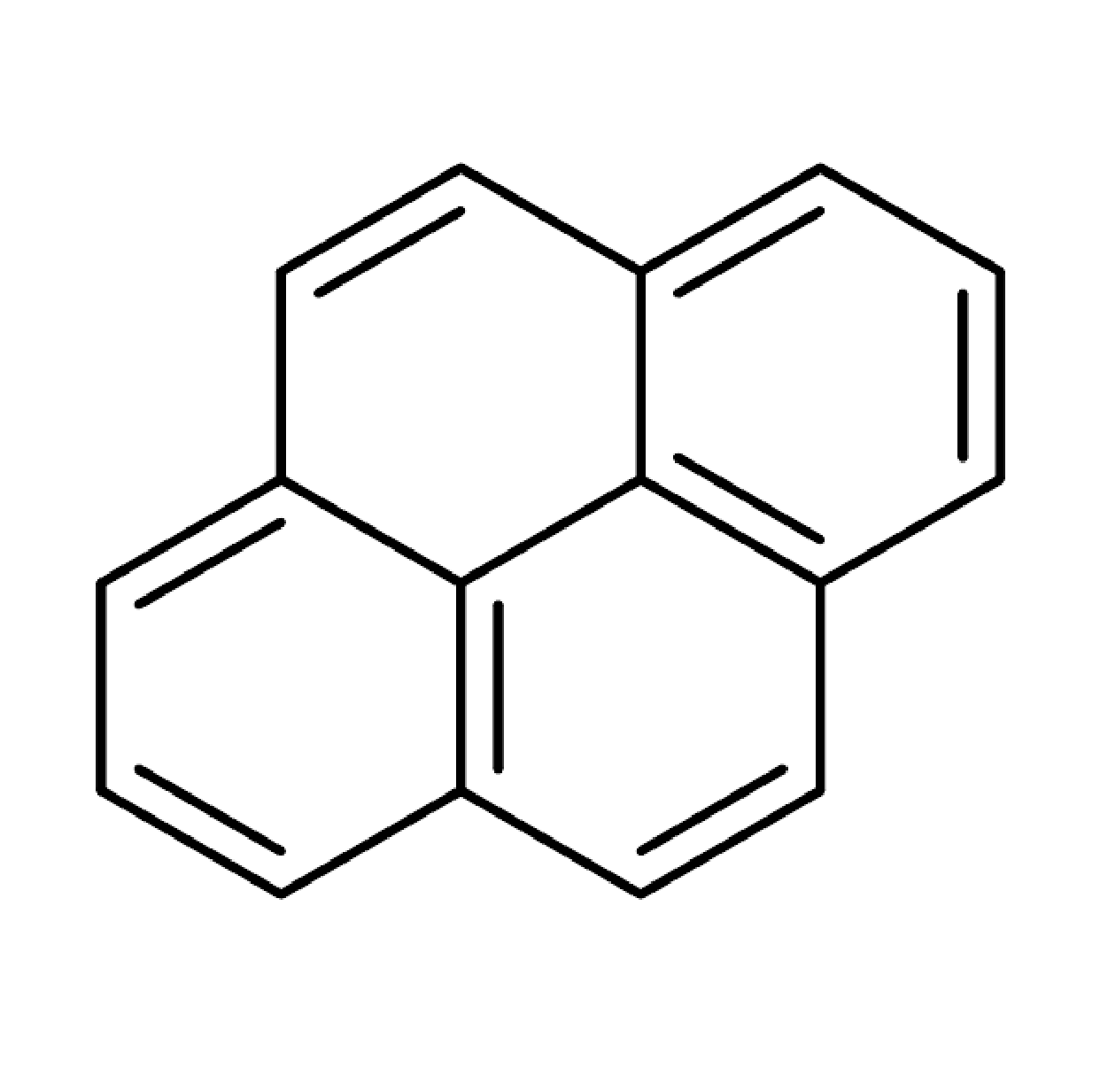}
				\includegraphics[scale=.17]{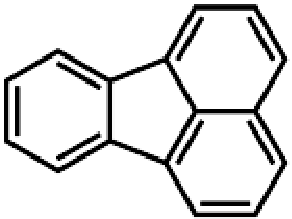}
			\end{subfigure}
			\begin{subfigure}[c]{0.94\textwidth}
				\includegraphics[width=\textwidth]{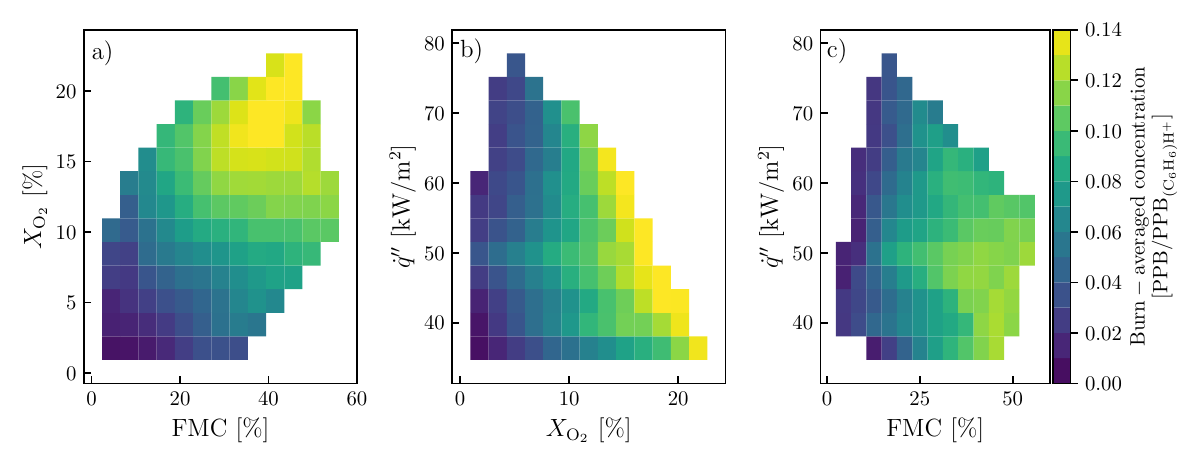}
			\end{subfigure}
			\caption{Averaged emission of fluoranthene/pyrene (FLA/PY) normalized to benzene as a function of the burn environment.
				Gas-phase emissions are reduced by 10.9\% if burn conditions are chosen in the optimal parameter range compared to conditions for sample C.}
			\label{fig:heat_maps_fluoranthene_pyrene}
		\end{figure*}

\begin{figure*}
			\centering
			\begin{subfigure}[c]{0.05\linewidth}
				\includegraphics[scale=.2]{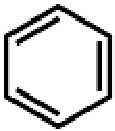}
			\end{subfigure}
			\begin{subfigure}[c]{0.94\textwidth}
				\includegraphics[width=\textwidth]{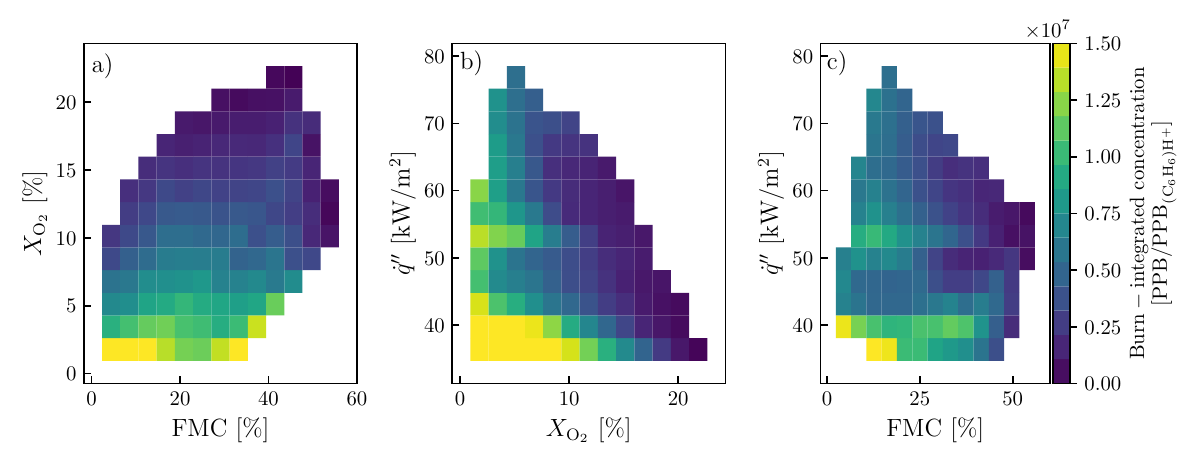}
			\end{subfigure}
			\caption{Averaged emission of benzene normalized by total sample mass loss.}
			\label{fig:heat_maps_benzene}
		\end{figure*}

Emission profiles for the remaining PAHs are shown in Figs.~\ref{fig:heat_maps_naphthalene} to \ref{fig:heat_maps_fluoranthene_pyrene}.
		They are obtained in the same manner as shown in Fig.~\ref{fig:big_picture} for PH/AN.
		Note that we deliberately excluded B[a]A/CHR and B[b]F/B[a]P due to low ion intensities in the gas phase.
		We also provide a non-normalized emission map for benzene ($\ce{C6H6)H+}$) in Fig.~\ref{fig:heat_maps_benzene} which shows gas-phase emissions normalized to the sample's mass loss (in units PPB/kg).
		
\end{document}